\documentclass[twocolumn,showpacs,superscriptaddress,amsmath,amssymb,aps,pra]{revtex4-1}
\usepackage{bm}
\usepackage{graphicx}
\usepackage{dcolumn}
\usepackage{amsmath,txfonts}
\usepackage{epstopdf}
\usepackage[mathlines]{lineno}
\usepackage{color}
\usepackage[colorlinks=true,linkcolor=blue,citecolor=blue]{hyperref}
\usepackage[normalem]{ulem}

\begin{document}

\title{Parity-symmetry-breaking quantum phase transition via parametric drive in a cavity magnonic system}

\author{Guo-Qiang Zhang}
\affiliation{Interdisciplinary Center of Quantum Information, State Key Laboratory of Modern Optical Instrumentation, and Zhejiang Province Key Laboratory of Quantum Technology and Device, Department of Physics, Zhejiang University, Hangzhou 310027, China}

\author{Zhen Chen}
\affiliation{Beijing Academy of Quantum Information Sciences, Beijing 100193, China}

\author{Wei Xiong}
\affiliation{Department of Physics, Wenzhou University, Zhejiang 325035, China}

\author{Chi-Hang Lam}
\affiliation{Department of Applied Physics, Hong Kong Polytechnic University, Hung Hom, Hong Kong, China}

\author{J. Q. You}
\email{jqyou@zju.edu.cn}
\affiliation{Interdisciplinary Center of Quantum Information, State Key Laboratory of Modern Optical Instrumentation, and Zhejiang Province Key Laboratory of Quantum Technology and Device, Department of Physics, Zhejiang University, Hangzhou 310027, China}

\begin{abstract}
We study the parity-symmetry-breaking quantum phase transition (QPT) in a cavity magnonic system driven by a parametric field, where the magnons in a ferrimagnetic yttrium-iron-garnet sphere strongly couple to a microwave cavity. With appropriate parameters, this cavity magnonic system can exhibit a rich phase diagram, including the parity-symmetric phase, parity-symmetry-broken phase, and bistable phase. When increasing the drive strength beyond a critical threshold, the cavity magnonic system undergoes either a first- or second-order nonequilibrium QPT from the parity-symmetric phase with microscopic excitations to the parity-symmetry-broken phase with macroscopic excitations, depending on the parameters of the system. Our work provides an alternate way to engineer the QPT in a hybrid quantum system containing the spin ensemble in a ferri- or ferromagnetic material with strong exchange interactions.
\end{abstract}

\date{\today}

\maketitle

\section{Introduction}

The light-matter interaction lies at the core of quantum technologies and has been a topic of lasting interest~\cite{Xiang13,Forn-Diaz19,Kockum2019}. A class of light-matter interactions, namely, the collective coupling between an ensemble of two-level systems and a single bosonic mode, can be described by the standard Dicke (or Tavis-Cummings) model~\cite{Dicke54}, equivalent to the model of two {\it nonlinearly coupled} harmonic oscillators~\cite{Emary-2003-PRE,Emary-2003-PRL}. When increasing the coupling strength to exceed a critical threshold, the Dicke model undergoes an equilibrium super-radiant quantum phase transition (QPT)~\cite{Emary-2003-PRE,Emary-2003-PRL,Hepp73,Wang73}. However, it is challenging to experimentally achieve this equilibrium super-radiant QPT due to the very large critical coupling strength. To circumvent this difficulty, an effective Dicke model has been designed or engineered with the assistance of external drive fields to study a nonequilibrium super-radiant QPT, both theoretically~\cite{Dimer07,Nagy10,You20,Zhu20} and experimentally~\cite{Baumann10,Baden14}, where the critical coupling strength is significantly reduced.

Among various systems composed of an ensemble of two-level systems interacting with a single bosonic mode, the cavity magnonic system has attracted much recent attention~\cite{Tabuchi14,Zhang14,SoykalPRL10,Zhang15-1,Cao15,Bai15,Tabuchi15,Li18,Shim20}, where magnons in, e.g., a single-crystal sphere of yttrium iron garnet (YIG), are strongly coupled to a microwave cavity mode~\cite{Lachance-Quirion19,Rameshti2021}. With flexible controllability and easily-engineered strong interactions, cavity magnonic systems have become a good platform for exploring various intriguing phenomena, such as magnon dark modes memory~\cite{Zhang15-2}, exceptional point~\cite{Zhang17,Zhang19,Zhao20,You19,Cao19}, dissipative magnon-photon coupling~\cite{Harder18,Grigoryan18,Wang19,Yu19} and magnon-induced nearly perfect absorption~\cite{Rao21}. Very recently, approaches to implement unconventional magnon excitations~\cite{Yuan-20-1} and stationary one-way quantum steering~\cite{Yang21} by driving the cavity magnonic system with a quantum squeezed field (i.e., via a parametric drive) were proposed.

In addition, magnon Kerr effect, i.e., a nonlinear effect due to the magnetocrystalline anisotropy in the YIG~\cite{Zhang-China-19}, was experimentally demonstrated~\cite{Wang16}. By pumping the cavity magnonic system with a coherent microwave field, bistability of cavity magnon polaritons was observed~\cite{Wang18} and tristability was subsequently predicted~\cite{Nair20,Bi21}. Moreover, based on the magnon Kerr effect, nonreciprocal transmission of a microwave field~\cite{Kong19} and the quantum entanglement between two magnon modes~\cite{Scully19} can be engineered in a ternary cavity magnonic system. However, due to the strong exchange interaction between nearest-neighbor spins, spin ensemble in the ferrimagnetic YIG, even under a strong microwave drive, remains in the low-lying excitations~\cite{Gurevich96,Stancil09}, and thus the magnon mode becomes {\it linearly coupled} to the cavity mode (i.e., the coupling strength between magnon and cavity modes is nearly independent of the magnon occupation)~\cite{Wang16,Wang18}. This hinders the occurrence of the QPT in this cavity magnonic system embedding the spin ensemble in the ferrimagnetic YIG, because two linearly-coupled bosonic modes do not exhibit any QPT. However, as shown below, we find that by taking account of the magnon Kerr effect in the YIG, the cavity magnonic system can resume the nonequilibrium QPT by driving the system with a squeezed microwave field. This provides a way to engineer the QPT in hybrid quantum systems containing spin ensembles in ferri- and ferromagnetic materials with strong exchange interactions.

In the rotating-reference frame with respect to the frequency of the drive field, the parametrically driven cavity magnonic system has an effective Hamiltonian that preserves the parity symmetry of the system in the absence of the driving. This is in sharp contrast to the cavity magnonic system with a coherent microwave driving~\cite{Wang16,Wang18}, in which the parity symmetry is not preserved. As a result, the parametrically driven cavity magnonic system can have a rich phase diagram, including the parity-symmetric phase, parity-symmetry-broken phase, and bistable phase. It can undergo a nonequilibrium parity-symmetry-breaking QPT at a critical drive strength (i.e., from a parity-symmetric phase with microscopic excitations to a parity-symmetry-broken phase with macroscopic excitations). Moreover, we find that this transition can be either discontinuous (first-order) or continuous (second-order), depending on the parameters of the system.

Our work provides an experimentally-realizable scheme to engineer QPT in a cavity magnonic system. In many-body quantum systems, the QPT is of fundamental importance in, e.g., understanding the semiclassical-to-quantum boundary~\cite{Sachdev11}. Moreover, owing to its close relation to quantum entanglement, the QPT may have potential applications in quantum information processing~\cite{Lambert04,Vidal03,Osborne02}. To experimentally demonstrate the QPT, it requires a well-controlled quantum system with tunable parameters. The cavity magnonic system offers an ideal platform for this, because it has flexible controllability in magnon frequency and coupling strength, as well as a tunable frequency and power of the drive field~\cite{Zhang17,Zhang19,Zhao20}.

The paper is organized as follows. In Sec.~\ref{model}, we give the Hamiltonian of the proposed system and then derive the steady-state solutions of the cavity magnonic system via a Heisenberg-Langevin approach. In Sec.~\ref{QPT}, by carrying out standard stability analysis, the QPT behaviors of the cavity magnonic system are shown in detail. Finally, discussions and conclusions are given in Sec.~\ref{conclusions}.

\begin{figure}
\includegraphics[width=0.42\textwidth]{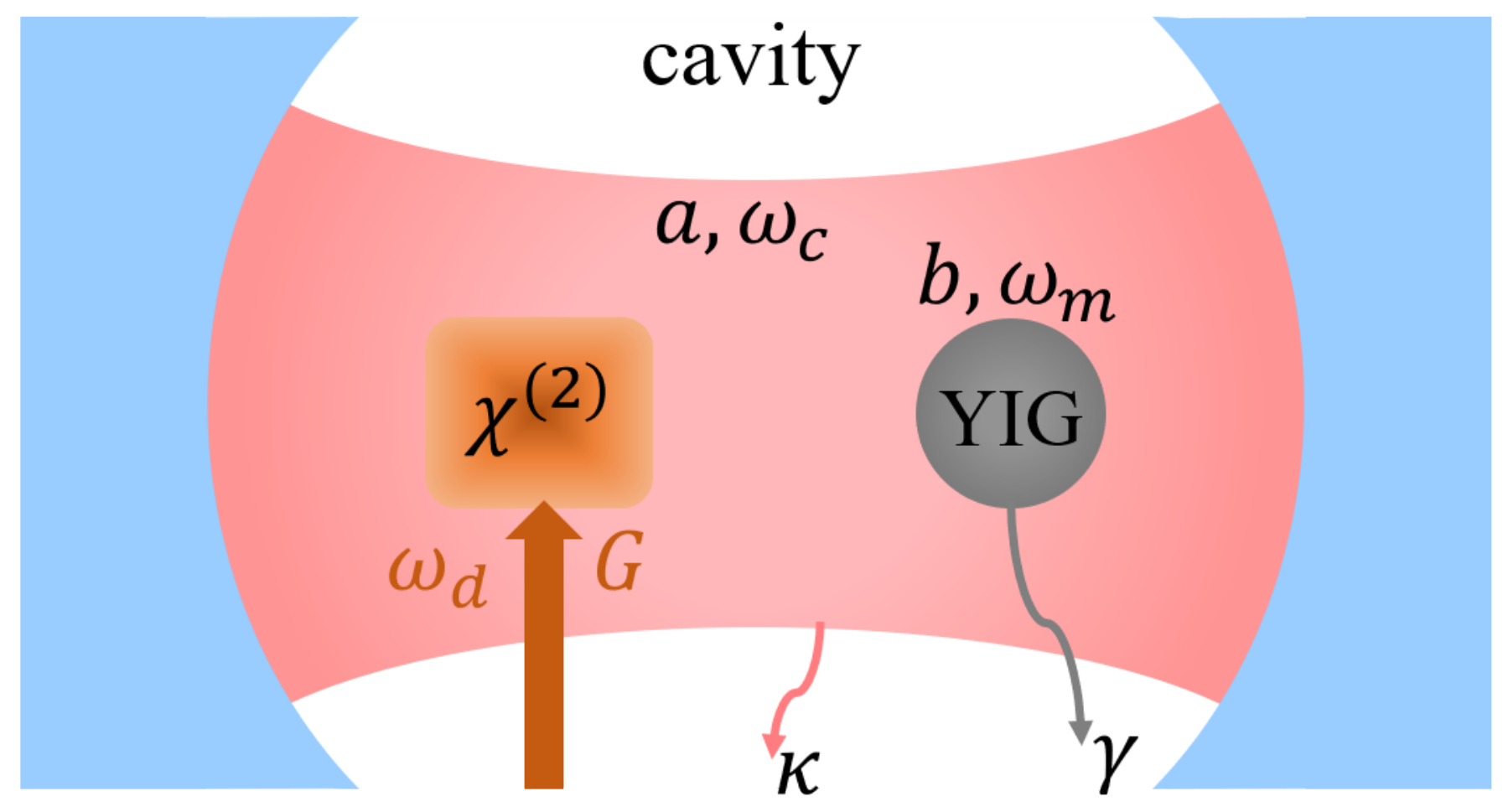}
\caption{Schematic of the proposed cavity magnonic system consisting of a small YIG sphere and a microwave cavity. The parametric drive on the cavity, with frequency $\omega_{d}$ and amplitude $G$, is generated by pumping a $\chi^{(2)}$ nonlinear medium inside the cavity. Here, $a$ is the annihilation operator of the cavity mode with frequency $\omega_{c}$ and decay rate $\kappa$, and $b$ is the annihilation operator of the magnon mode with frequency $\omega_{m}$ and damping rate $\gamma$.}
\label{fig1}
\end{figure}

\section{The Model}\label{model}

The cavity magnonic system, depicted in Fig.~\ref{fig1}, consists of a small YIG sphere and a parametrically driven microwave cavity. Here we focus on the Kittel mode of the spin waves in the YIG sphere, with all exchange-coupled spins precessing in phase together. Without considering the nonlinearity of magnons and the parametric drive on the cavity mode, the Hamiltonian of the proposed hybrid system reads (setting $\hbar=1$)
\begin{eqnarray}\label{main-Hamiltonian}
H_{s}=\omega_{c}a^{\dag}a+\omega_{m}b^{\dag}b+g(a^{\dag}b+ab^{\dag}),
\end{eqnarray}
where $a$ and $a^{\dag}$ ($b$ and $b^{\dag}$) are the annihilation and creation operators, respectively, of the cavity (magnon) mode with frequency $\omega_{c}$ ($\omega_{m}$), and $g$ is the coupling strength between the cavity and magnon modes. In the YIG sphere, the magnetocrystalline anisotropy gives rise to the interaction among magnons, described by the Kerr-nonlinearity Hamiltonian
\begin{eqnarray}\label{Kerr}
H_{\rm Kerr}=\frac{K}{2}b^{\dag}b^{\dag}bb.
\end{eqnarray}
The Kerr coefficient is positive (i.e., $K>0$) when the crystallographic axis [100] of the YIG sphere is parallel to the external static magnetic field~\cite{Wang18}. The microwave cavity contains a $\chi^{(2)}$ nonlinear medium pumped by an external field with frequency $\omega_{d}$. The corresponding Hamiltonian is
\begin{eqnarray}\label{parametric-drive}
H_{d}=\frac{G}{2}(a^{\dag}a^{\dag}e^{-i\omega_{d}t}+aae^{i\omega_{d}t}),
\end{eqnarray}
where the strength $G$ of the parametric drive can be tuned by varying the amplitude of the drive field. In a frame rotating with $\omega_{d}/2$, the total Hamiltonian of the system, $H_{\rm tot}=H_{s}+H_{\rm Kerr}+H_{d}$, can be written as
\begin{eqnarray}\label{Hermitian-Hamiltonian}
H_{\rm tot}&=&\Delta_{c}a^{\dag}a+\Delta_{m}b^{\dag}b+\frac{K}{2}b^{\dag}b^{\dag}bb+g(a^{\dag}b+ab^{\dag})\nonumber\\
           & &+\frac{G}{2}(a^{\dag}a^{\dag}+aa),
\end{eqnarray}
where $\Delta_{c(m)}=\omega_{c(m)}-\omega_{d}/2~(>0)$ is the frequency detuning of the cavity (magnon) mode relative to the drive field. Without the YIG sphere (i.e., $g=0$), the parametrically driven cavity can become unstable when $G>\sqrt{\Delta_{c}^{2}+\kappa^{2}}$~\cite{Walls94}, with $\kappa$ being the decay rate of the cavity mode. Hereafter, we only focus on the case of $K  > 0$ and $G < \sqrt{\Delta_{c}^{2}+\kappa^{2}}$.

With the Hamiltonian in Eq.~(\ref{Hermitian-Hamiltonian}), the dynamics of the cavity magnonic system can be described by the quantum Langevin equations~\cite{Walls94}:
\begin{eqnarray}\label{Langevin}
\dot{a}&=&-i(\Delta_{c}-i\kappa)a-igb-iGa^{\dag}+\sqrt{2\kappa}a_{\rm{in}},\nonumber\\
\dot{b}&=&-i(\Delta_{m}-i\gamma)b-iKb^{\dag}bb-iga+\sqrt{2\gamma}b_{\rm{in}},
\end{eqnarray}
where $\gamma$ is the damping rate of the magnon mode, and $a_{\rm{in}}$ ($b_{\rm{in}}$) is the noise operator for the cavity (magnon) mode. These noise operators have zero mean values, i.e., $\langle a_{\rm{in}}\rangle=\langle b_{\rm{in}}\rangle=0$. Decoherence can arise from both energy relaxation (i.e., dissipation) and pure dephasing. In the cavity magnonic system, dephasing effects of the cavity and magnon modes can be ignored (cf. Appendix~\ref{Appendix-A}) and only dissipations are considered in Eq.~(\ref{Langevin}).

The equations of motion in Eq.~(\ref{Langevin}) can be rewritten as $\dot{a}=-i[a, H_{\rm eff}]+\sqrt{2\kappa}a_{\rm{in}}$ and $\dot{b}=-i[b, H_{\rm eff}]+\sqrt{2\gamma}b_{\rm{in}}$, where
\begin{eqnarray}\label{non-Hermitian-Hamiltonian}
H_{\rm eff}&=&(\Delta_{c}-i\kappa)a^{\dag}a+(\Delta_{m}-i\gamma)b^{\dag}b+\frac{K}{2}b^{\dag}b^{\dag}bb\nonumber\\
              && +g(a^{\dag}b+ab^{\dag})+\frac{G}{2}(a^{\dag}a^{\dag}+aa)
\end{eqnarray}
is the effective non-Hermitian Hamiltonian of the cavity magnonic system. In Eq.~(\ref{non-Hermitian-Hamiltonian}), $-i\kappa a^{\dag}a$ and $-i\gamma b^{\dag}b$ describe the dissipations of the cavity and magnon modes, respectively. Note that the drive term proportional to $a^{\dag}a^{\dag}$ ($aa$) corresponds to the simultaneous creation (destruction) of two photons in the cavity, which does not preserve the total number of excitations in the system. However, the parity symmetry is preserved in the non-Hermitian Hamiltonian (\ref{non-Hermitian-Hamiltonian}), $[H_{\rm eff},\Pi]=0$, where $\Pi=\exp[i\pi(a^{\dag}a+b^{\dag}b)]$ is the parity operator~\cite{Emary-2003-PRE,Emary-2003-PRL}.

In our scheme, both the magnon Kerr effect and the parametric drive play the key roles in engineering the parity-symmetry-breaking QPT (cf.~Sec.~\ref{QPT}). Usually, the QPT results from the nonlinear interaction of the system. For example, in the standard Dicke model (equivalent to two {\it nonlinearly coupled} harmonic oscillators), the super-radiant QPT is induced by the nonlinear coupling~\cite{Emary-2003-PRE,Emary-2003-PRL}, which plays a similar role as the magnon Kerr effect in our proposed system. As shown in Ref.~\cite{Yuan-20-1}, the cavity magnonic system with a parametric drive can exhibit unconventional magnon excitations but no QPT occurs in the case without magnon Kerr effect, which corresponds to $K=0$ in Eq.~(\ref{non-Hermitian-Hamiltonian}). On the other hand, if the cavity or the YIG sphere is pumped by a coherent field instead of a squeezed field [corresponding to Eq.~(\ref{non-Hermitian-Hamiltonian}) with $a^{\dag}a^{\dag}+aa$ replaced by $a^{\dag}+a$ or $b^{\dag}+b$], the coherent drive will break the parity symmetry of the cavity magnonic system, i.e., $[H_{\rm eff},\Pi] \neq 0$. In such a case, the cavity magnonic system exhibits the nonlinear foldover effect rather than the QPT in the steady state, which was demonstrated experimentally in Refs.~\cite{Wang18,Hyde18}.

To analyze the steady-state properties of the system, we write the operator $a$ ($b$) as the sum of the expectation value and its fluctuation: $a=\langle a\rangle+\delta a$, and $b=\langle b\rangle+\delta b$, where $\langle \delta a\rangle=\langle \delta b\rangle=0$. From Eq.~(\ref{Langevin}), it follows that
\begin{eqnarray}\label{expected-value}
\langle\dot{a}\rangle&=&-i(\Delta_{c}-i\kappa)\langle a\rangle -ig\langle b\rangle-iG\langle a^{\dag}\rangle,\nonumber\\
\langle\dot{b}\rangle&=&-i(\Delta_{m}+K\langle b^{\dag}\rangle\langle b\rangle-i\gamma)\langle b\rangle-ig\langle a\rangle,
\end{eqnarray}
and
\begin{eqnarray}\label{fluctuations}
\delta\dot{a}&=&-i(\Delta_{c}-i\kappa)\delta a-ig\delta b-iG\delta a^{\dag}+\sqrt{2\kappa}a_{\rm{in}},\nonumber\\
\delta\dot{b}&=&-i(\widetilde{\Delta}_{m}-i\gamma)\delta b-ig\delta a-iF\delta b^{\dag}+\sqrt{2\gamma}b_{\rm{in}},
\end{eqnarray}
with $\widetilde{\Delta}_{m}=\Delta_{m}+2K\langle b^{\dag}\rangle\langle b\rangle$ and $F=K\langle b\rangle^{2}$, where the high-order terms of the fluctuations have been neglected. As shown in Appendix~\ref{Appendix-A}, the equations of motion for the expectation values $\langle a\rangle$ and $\langle b\rangle$ in Eq.~(\ref{expected-value}) can also be derived via a master equation approach~\cite{Walls94}. This mean-field approach can give accurate results in the thermodynamic limit $\gamma/K \rightarrow +\infty$~\cite{Zhang20}. Indeed, this is the case in our cavity magnonic system, because $K \ll \gamma$~\cite{Zhang-China-19,Wang18}.

\begin{figure}
\includegraphics[width=0.48\textwidth]{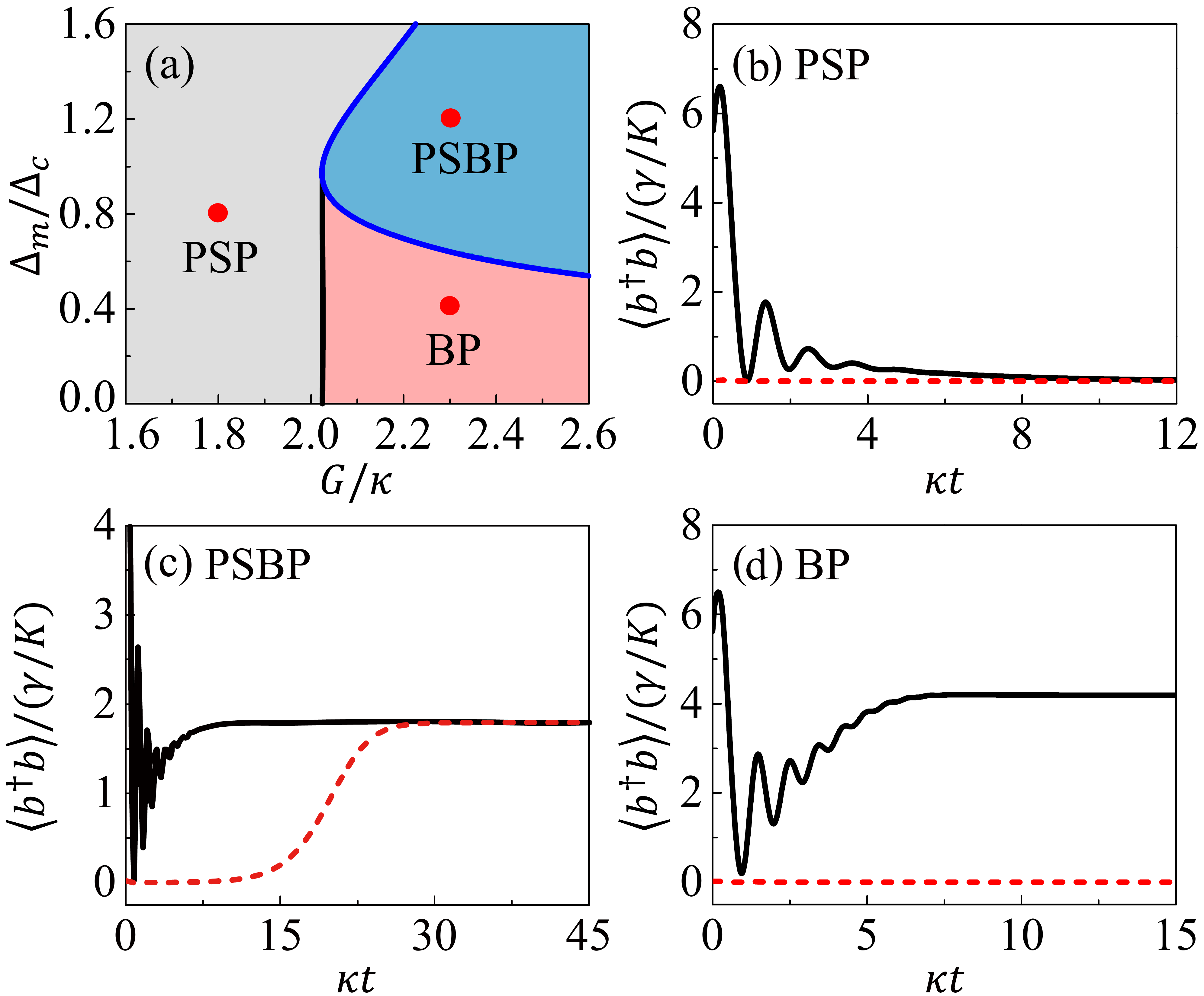}
\caption{(a) Steady-state phase diagram of the cavity magnonic system, where PSP, PSBP and BP denote the parity-symmetric phase, parity-symmetry-broken phase and bistable phase, respectively. (b)-(d) Time evolution of the scaled magnon number $\langle b^{\dag}b\rangle/(\gamma/K)$, obtained by numerically solving Eq.~(\ref{expected-value}) at the points indicated by red dots in (a), with (b) $G/\kappa=1.8$, and $\Delta_{m}/\Delta_{c}=0.8$, (c) $G/\kappa=2.3$, and $\Delta_{m}/\Delta_{c}=1.2$, and (d) $G/\kappa=2.3$, and $\Delta_{m}/\Delta_{c}=0.4$. In (b)-(d), $\langle a\rangle_{t=0}/\sqrt{\gamma/K}=2.1+1.1i$, and $\langle b\rangle_{t=0}/\sqrt{\gamma/K}=2.1-1.1i$ for the (black) solid curves, while $\langle a\rangle_{t=0}/\sqrt{\gamma/K}=0.1+0.1i$, and $\langle b\rangle_{t=0}/\sqrt{\gamma/K}=0.1-0.1i$ for the (red) dashed curves. Other parameters are $\Delta_{c}/\kappa=3$, $g/\kappa=2.4$, and $\gamma/\kappa=1$.}
\label{fig2}
\end{figure}

At the steady state, i.e., $\langle\dot{a}\rangle=\langle\dot{b}\rangle=0$ in Eq.~(\ref{expected-value}), we obtain
one trivial solution $\langle b^{\dag}b\rangle_{0}$ and two nontrivial solutions $\langle b^{\dag}b\rangle_{\pm}$:
\begin{equation}\label{solutions}
\langle b^{\dag}b\rangle_{0}=0,~~~
\langle b^{\dag}b\rangle_{\pm}=\Big(-\Delta'_{m} \pm \sqrt{\eta^{2}G^{2}-{\gamma'}^{2}}\Big)/K,
\end{equation}
where $\Delta'_{m}=\Delta_{m}-\eta\Delta_{c}$, $\gamma'= \gamma+\eta\kappa$, and $\eta=g^{2}/(\Delta_{c}^{2}+\kappa^{2}-G^{2})$. Also, the mean-field approximation $\langle b^{\dag}b\rangle \approx \langle b^{\dag}\rangle\langle b\rangle$ has been used. Obviously, the drive strength $G$ should be in an appropriate regime to ensure $\langle b^{\dag}b\rangle_{+} > 0$. Note that because another nontrivial solution $\langle b^{\dag}b\rangle_{-}$ is unstable [see Fig.~\ref{fig2}(a) and related discussions], we need not consider it here.

Solving $\langle b^{\dag}b\rangle_{+} > 0$, we obtain $G>G_{\rm c1}$ in the case of $\Delta_{m}/\Delta_{c} < \zeta$, with
\begin{equation}\label{zeta}
\zeta=\frac{2\gamma^{2}}{\sqrt{4(\Delta_{c}^{2}+\kappa^{2})\gamma^{2}+(4\kappa\gamma+g^{2})g^{2}}-(2\kappa\gamma+g^{2})},
\end{equation}
where the critical drive strength is
\begin{equation}\label{Gc1}
G_{\rm c1}=\sqrt{\Delta_{c}^{2}+\kappa^{2}+(4\kappa\gamma+g^{2})g^{2}/4\gamma^{2}}-g^{2}/2\gamma.
\end{equation}
However, when $\Delta_{m}/\Delta_{c} > \zeta$, to achieve the valid solution $\langle b^{\dag}b\rangle_{+}>0$, the constraint on the drive strength $G$ becomes $G>G_{\rm c2}$, where
\begin{eqnarray}\label{Gc2}
G_{\rm c2}=\sqrt{\Delta_{c}^{2}+\kappa^{2}-(2\Delta_{c}\Delta_{m}-2\kappa\gamma-g^{2})g^{2}/(\Delta_{m}^{2}+\gamma^{2})}~~~
\end{eqnarray}
corresponds to another critical drive strength.
Here, $G_{\rm c1}$ is independent of $\Delta_{m}$, while $G_{\rm c2}$ depends on $\Delta_{m}$.  In particular, $G_{\rm c1}=G_{\rm c2}$ at $\Delta_{m}/\Delta_{c} = \zeta$.
From the second equation in Eq.~(\ref{expected-value}), when $\langle\dot{b}\rangle=0$, it follows that the steady-state photon occupation can be expressed as
\begin{eqnarray}\label{}
\langle a^{\dag}a\rangle=\big[(\Delta_{m}+K\langle b^{\dag}b\rangle)^{2}+\gamma^{2}\big]\langle b^{\dag}b\rangle/g^{2}.
\end{eqnarray}
Therefore, we need only to focus on the magnon occupation $\langle b^{\dag}b\rangle$, which can be defined as an order parameter of our hybrid system.

\section{QPT in the cavity magnonic system}\label{QPT}

To characterize the steady-state phase diagram of the cavity magnonic system, we analyze the stability of the solutions for $\langle b^{\dag}b\rangle$ in Eq.~(\ref{solutions}). Defining a column vector $\delta\mathbf{v}=(\delta a,\delta b,\delta a^{\dag},\delta b^{\dag})^{T}$ of the fluctuations, we can write Eq.~(\ref{fluctuations}) and its complex conjugate in a matrix form,
\begin{equation}
\delta\dot{\mathbf{v}}=\mathbf{M}\delta\mathbf{v}+\delta\mathbf{v}_{\rm in},
\end{equation}
where $\delta\mathbf{v}_{\rm in}=(\sqrt{2\kappa}a_{\rm{in}},\sqrt{2\gamma}b_{\rm{in}},
\sqrt{2\kappa}a_{\rm{in}}^{\dag},\sqrt{2\gamma}b_{\rm{in}}^{\dag})^{T}$ is the column vector of the noise operators, and
\begin{equation}\label{matrix}
\mathbf{M}=
\left(
  \begin{array}{cccc}
    -i\Delta_{c}-\kappa& -ig & -iG&0\\
    -ig & -i\widetilde{\Delta}_{m}-\gamma & 0&-iF\\
    iG & 0 & i\Delta_{c}-\kappa  &ig\\
    0 & iF^{*} & ig&i\widetilde{\Delta}_{m}-\gamma\\
  \end{array}
\right).
\end{equation}
For a given solution of $\langle b^{\dag}b\rangle$ in Eq.~(\ref{solutions}), it is stable only if the real parts of all eigenvalues of the matrix $\mathbf{M}$ are negative~\cite{Gradshteyn80}.

\begin{figure}
\includegraphics[width=0.43\textwidth]{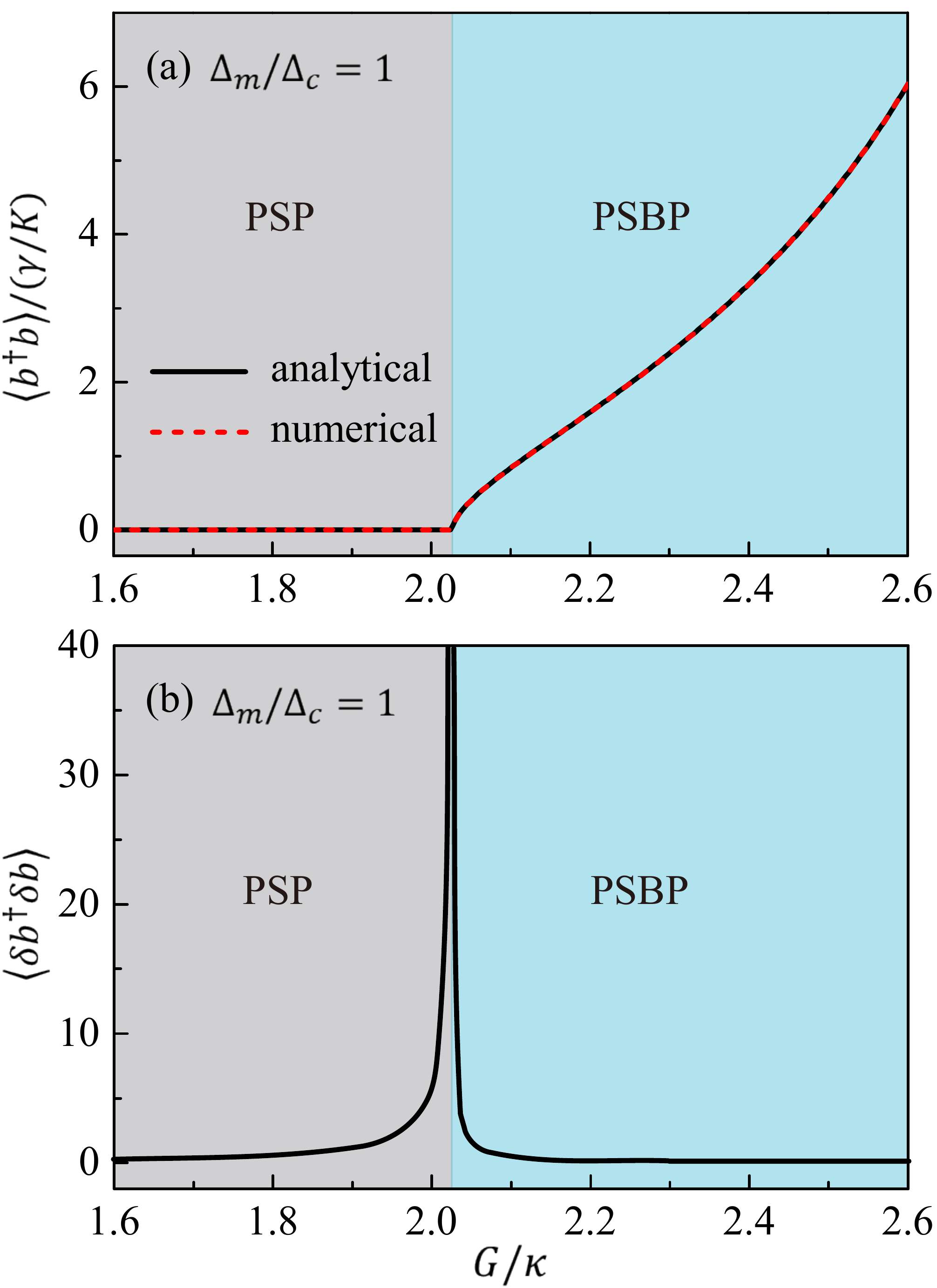}
\caption{(a) The scaled steady-state magnon number $\langle b^{\dag}b\rangle/(\gamma/K)$ and (b) the expectation value $\langle\delta b^{\dag}\delta b\rangle$ versus the reduced drive strength $G/\kappa$ in the case of $\Delta_{m}/\Delta_{c} = 1$. In (a), the (black) solid curve corresponds to the analytical results in Eq.~(\ref{solutions}), and the (red) dashed curve corresponds to the numerical results obtained using Eq.~(\ref{expected-value}) with initial conditions $\langle a\rangle_{t=0}/\sqrt{\gamma/K}=\langle b\rangle_{t=0}/\sqrt{\gamma/K}=0.001$. Other parameters are the same as in Fig.~\ref{fig2}(a).}
\label{fig3}
\end{figure}

\begin{figure}
\includegraphics[width=0.43\textwidth]{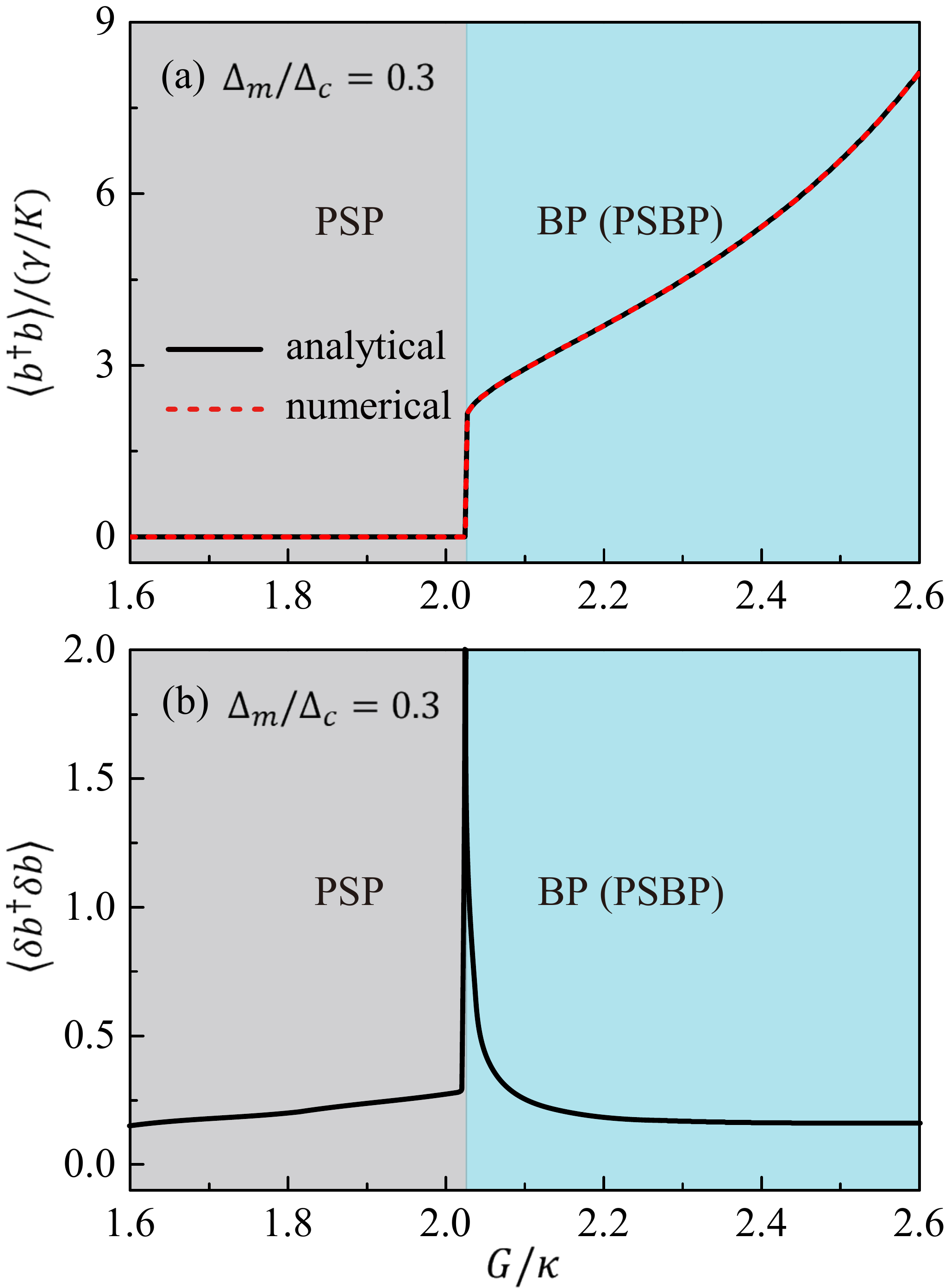}
\caption{(a) The scaled steady-state magnon number $\langle b^{\dag}b\rangle/(\gamma/K)$ and (b) the expectation value $\langle\delta b^{\dag}\delta b\rangle$ versus the reduced drive strength $G/\kappa$ in the case of $\Delta_{m}/\Delta_{c} = 0.3$, where BP (PSBP) denotes the parity-symmetry-broken phase in the bistable region. In (a), the (black) solid curve corresponds to the analytical results in Eq.~(\ref{solutions}), and the (red) dashed curve corresponds to the numerical results obtained using Eq.~(\ref{auxiliary-pulse}) with $\Omega_{0}/\kappa\sqrt{\gamma/K}=2$, $\kappa\tau=10$, and the initial conditions $\langle a\rangle_{t=0}/\sqrt{\gamma/K}=\langle b\rangle_{t=0}/\sqrt{\gamma/K}=0$. Other parameters are the same as in Fig.~\ref{fig2}(a).}
\label{fig4}
\end{figure}

By carrying out the standard stability analysis, we display in Fig.~\ref{fig2}(a) the steady-state phase diagram versus the reduced drive strength $G/\kappa$ and the ratio $\Delta_{m}/\Delta_{c}$ of the frequency detunings of the cavity and magnon modes. There are three different regions in the phase diagram, i.e., the parity-symmetric phase (gray area), parity-symmetry-broken phase (blue area), and bistable phase (red area), where the vertical black solid line ($G=G_{\rm c1}$) and the blue solid curve ($G=G_{\rm c2}$) are the boundaries between different phases. In the parity-symmetric phase, there is no macroscopic magnon excitations and the solution $\langle b^{\dag}b\rangle=0$ is stable [see Fig.~\ref{fig2}(b)]. The nontrivial solution $\langle b^{\dag}b\rangle=\langle b^{\dag}b\rangle_{+}$ given in Eq.~(\ref{solutions}) becomes stable in the parity-symmetry-broken phase [see Fig.~\ref{fig2}(c)], which has macroscopic  magnon excitations (i.e., $\langle b^{\dag}b\rangle\ne 0$). In the bistable phase, both solutions $\langle b^{\dag}b\rangle=0$ and $\langle b^{\dag}b\rangle=\langle b^{\dag}b\rangle_{+}$ are stable, and the initial condition determines the steady-state magnon occupation [see Fig.~\ref{fig2}(d)]. Moreover, we find that the nontrivial solution $\langle b^{\dag}b\rangle=\langle b^{\dag}b\rangle_{-}$ is unstable in the whole parameter space, since the real parts of the four eigenvalues of the related matrix $\mathbf{M}$ in Eq.~(\ref{matrix}) are not all negative.
With the parameters of the system used in Fig.~\ref{fig2}(a), the coefficient $\zeta$ given in Eq.~(\ref{zeta}) is $\zeta=0.976$.
When $\Delta_{m}/\Delta_{c}>\zeta$ ($\Delta_{m}/\Delta_{c}<\zeta$), the boundary between parity-symmetric and parity-symmetry-broken (bistable) phases is given by $G=G_{\rm c2}$ ($G=G_{\rm c1}$).

In the case of $\Delta_{m}/\Delta_{c}>\zeta$ (e.g., $\Delta_{m}/\Delta_{c}=1$), we plot the scaled steady-state magnon number $\langle b^{\dag}b\rangle/(\gamma/K)$ versus the reduced drive strength $G/\kappa$ in Fig.~\ref{fig3}(a). Here the QPT occurs at the critical drive strength $G_{\rm c2}/\kappa=2.025$, where $\langle b^{\dag}b\rangle/(\gamma/K)$ changes continuously, indicating a {\it second-order} QPT. When $G < G_{\rm c2}$, the system is in the parity-symmetric phase with $\langle b^{\dag}b\rangle=0$. However, when $G > G_{\rm c2}$, it is in the parity-symmetry-broken phase with $\langle b^{\dag}b\rangle=\langle b^{\dag}b\rangle_{+}\ne 0$, as given in Eq.~(\ref{solutions}). Around the critical point $G=G_{\rm c2}$, it can be obtained that (see Appendix~\ref{Appendix-B})
\begin{equation}\label{critical-exponent}
\langle b^{\dag}b\rangle \sim |G-G_{\rm c2}|^{\nu},
\end{equation}
with the critical exponent $\nu=1$. To further study the nonequilibrium QPT, we also investigate the response of the expectation value of the correlated fluctuation $\delta b^{\dag}\delta b$. Here we only give the main numerical results. The equations of motion for this and other related correlated fluctuations can be found in Appendix~\ref{Appendix-C}. As shown in Fig.~\ref{fig3}(b), the expectation value $\langle \delta b^{\dag}\delta b\rangle$ approaches zero when the drive strength $G$ is tuned away from the critical value $G=G_{\rm c2}$, but diverges when $G$ is close to $G=G_{\rm c2}$~\cite{Nagy10,Nagy11,Zheng11}.

In Fig.~\ref{fig4}(a), we display the steady-state magnon number $\langle b^{\dag}b\rangle/(\gamma/K)$ versus the drive strength $G/\kappa$ in the case of $\Delta_{m}/\Delta_{c}<\zeta$ (e.g., $\Delta_{c}/\Delta_{m} = 0.3$). When varying the drive strength $G$ from $G<G_{\rm c1}$ to $G>G_{\rm c1}$, $\langle b^{\dag}b\rangle/(\gamma/K)$ is discontinuous at the critical point $G_{\rm c1}/\kappa=2.024$. This indicates that the cavity magnonic system undergoes a {\it first-order} QPT from the parity-symmetric phase with $\langle b^{\dag}b\rangle=0$ to the parity-symmetry-broken phase with $\langle b^{\dag}b\rangle=\langle b^{\dag}b\rangle_{+}\ne 0$, as given in Eq.~(\ref{solutions}). In such a case, the expectation value of the correlated fluctuation $\delta b^{\dag}\delta b$ is also divergent around the critical point $G=G_{\rm c1}$ [see Fig.~\ref{fig4}(b)].

\begin{figure}
\includegraphics[width=0.48\textwidth]{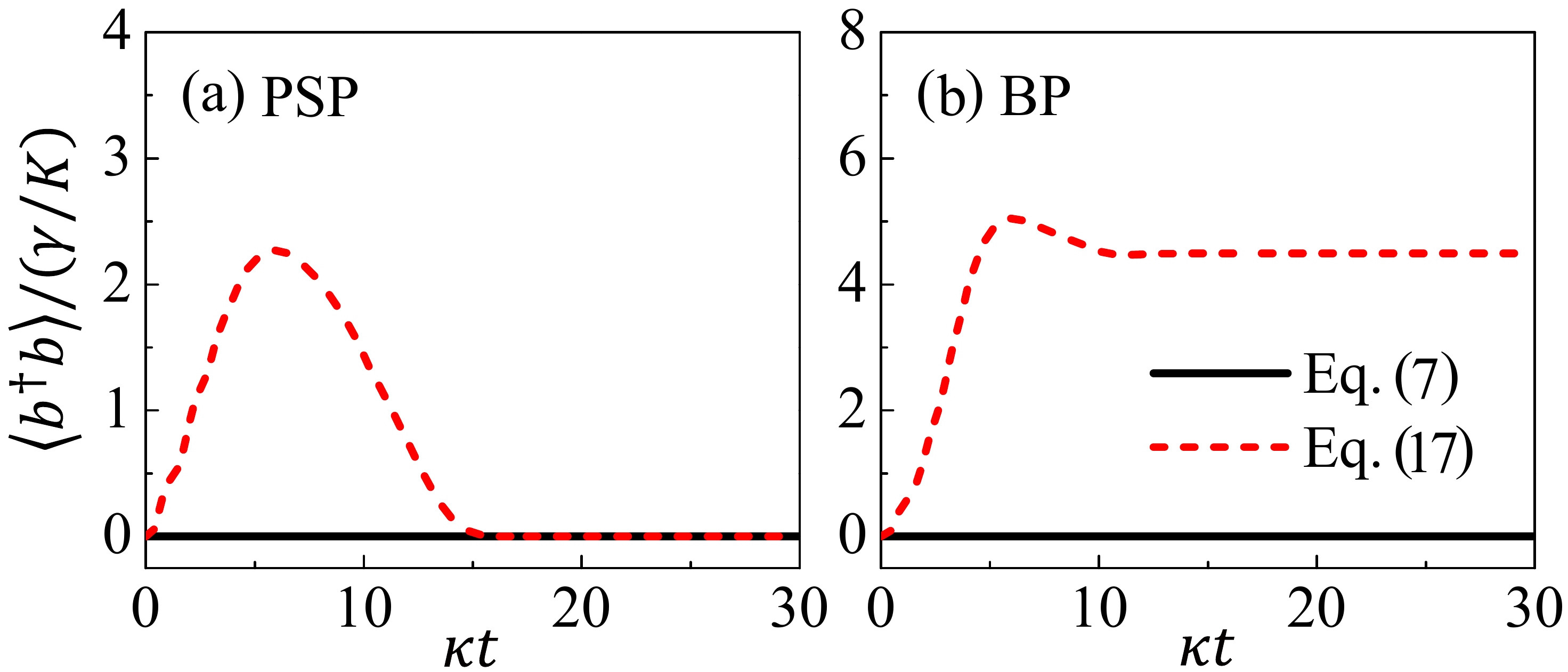}
\caption{Time evolution of the scaled magnon number $\langle b^{\dag}b\rangle/(\gamma/K)$ in the (a) parity-symmetric phase with $G/\kappa=1.8$ and (b) bistable phase with $G/\kappa=2.3$, where the (black) solid and (red) dashed curves correspond to the numerical results obtained using Eqs.~(\ref{expected-value}) and (\ref{auxiliary-pulse}), respectively. Other parameters are the same as in Fig.~\ref{fig4}(a).}
\label{fig5}
\end{figure}

Also, in Figs.~\ref{fig3}(a) and \ref{fig4}(a), we numerically simulate the QPT behaviors of the system (the red dashed curves), which agree well with the analytical results (the black solid curves). When $\Delta_{m}/\Delta_{c}>\zeta$, the numerical results related to the {\it second-order} QPT in Fig.~\ref{fig3}(a) can be easily obtained by solving Eq.~(\ref{expected-value}), where the initial conditions are slightly deviated from $\langle a\rangle_{t=0}/\sqrt{\gamma/K}=\langle b\rangle_{t=0}/\sqrt{\gamma/K}=0$. Different from the case of $\Delta_{m}/\Delta_{c}>\zeta$, when $\Delta_{m}/\Delta_{c}<\zeta$, the steady-state magnon occupation depends on the initial conditions in the bistable phase [see Fig.~\ref{fig2}(d)]. To engineer the {\it first-order} QPT in the experiment, one can use an auxiliary microwave pulse with frequency $\omega_{d}/2$ and duration $\tau$ to drive the cavity. This corresponds to adding the drive term $\Omega(t)(a^{\dag}+a)$ to the total Hamiltonian in Eq.~(\ref{Hermitian-Hamiltonian}), where the Rabi frequency $\Omega(t)=\Theta(\tau-t)\Omega_{0}\cos[\pi t/(2\tau)]$ describes the shape of the pulse, $\Omega_{0}$ is the Rabi frequency at $t=0$, and $\Theta(\tau-t)$ is the Heaviside function. When including the effect of the auxiliary microwave pulse, the dynamical equations of the expectation values $\langle a\rangle$ and $\langle b\rangle$ in Eq.~(\ref{expected-value}) becomes
\begin{eqnarray}\label{auxiliary-pulse}
\langle\dot{a}\rangle&=&-i(\Delta_{c}-i\kappa)\langle a\rangle -ig\langle b\rangle
                         -iG\langle a^{\dag}\rangle-i\Omega(t),\nonumber\\
\langle\dot{b}\rangle&=&-i(\Delta_{m}+K\langle b^{\dag}\rangle\langle b\rangle-i\gamma)\langle b\rangle-ig\langle a\rangle.
\end{eqnarray}
Note that in the region $t>\tau$, Eq.~(\ref{auxiliary-pulse}) is reduced to Eq.~(\ref{expected-value}) due to $\Omega(t)=0$. With appropriate values of the parameters $\Omega_{0}$ and $\tau$, the system evolves from the initial conditions $\langle a\rangle_{t=0}/\sqrt{\gamma/K}=\langle b\rangle_{t=0}/\sqrt{\gamma/K}=0$ to the parity-symmetry-broken phase instead of the parity-symmetric phase in the bistable region (see Fig.~\ref{fig5}). By solving Eq.~(\ref{auxiliary-pulse}), we obtain the numerical results of the {\it first-order} QPT in Fig.~\ref{fig4}(a) (the red dashed curves).

\section{Discussions and conclusions}\label{conclusions}

For a practical implementation of the nonequilibrium QPT, we can design a cavity magnonic system consisting of a small YIG sphere and a coplanar waveguide resonator (CWR). In the YIG sphere, the nonlinear interaction among magnons resulting from the magnetocrystalline anisotropy can be modeled by the Kerr-nonlinearity Hamiltonian~\cite{Zhang-China-19,Wang16,Wang18}. Experimentally, strong coupling of magnons to a given mode of the CWR has been demonstrated at tens of milli-Kelvin temperatures~\cite{Huebl13,Morris17,Hou19,Li19}. With the magnon Kerr effect considered, the cavity magnonic system can be equivalently described as a harmonic oscillator linearly coupled to a Kerr-nonlinearity oscillator. Typically, the decay rate $\kappa$ of the CWR mode, the damping rate $\gamma$ of the magnon mode, and the coupling strength $g$ between the magnon and CWR modes can be taken as~\cite{Morris17} $\kappa/2\pi=2.04$~MHz, $\gamma/2\pi=1.49$~MHz, and $g/2\pi=8.17$~MHz, respectively. In the parity-symmetry-broken phase, the magnon occupation is of the order $\gamma/K\sim 10^{16}$ [see Eq.~(\ref{solutions})], which is much smaller than the number of spins in the YIG sphere ($\sim 10^{19}$)~\cite{Zhang-China-19,Wang16,Wang18}. Therefore, only low-lying excitations occur for the spin ensemble in the YIG sphere. With the decay rate $\kappa/2\pi=2.04$~MHz, the corresponding critical strength of the parametric drive is $G/2\pi \approx 4.1$~MHz [cf. Figs.~\ref{fig3}(a) and \ref{fig4}(a)]. The parametric drive on the CWR can be provided by a flux-driven Josephson parametric amplifier with a pump field at frequency $\omega_{d}$~\cite{Yamamoto08,Zhong13,Lin13}, where the amplitude $G$ of the parametric drive can be controlled (ranging from 0 to 6 MHz) by the strength of the flux drive~\cite{Krantz16}. Because the flux-driven Josephson parametric amplifier is spatially separated from the YIG sphere, the parametric drive on the magnon mode can be screened. When including the dissipations of the system and the parametric drive, the hybrid system can be described by the effective non-Hermitian Hamiltonian in Eq.~(\ref{non-Hermitian-Hamiltonian}). Then, with appropriate parameters, the designed cavity magnonic system can display the parity-symmetry-breaking QPT by varying the strength $G$ of the parametric drive [see Figs.~\ref{fig3}(a) and \ref{fig4}(a)].

For an ideal flux-driven Josephson parametric amplifier, there is no coherent drive on the cavity~\cite{Yamamoto08,Zhong13,Lin13}. However, in the experiment, the flux-driven Josephson parametric amplifier is not so ideal that an additional coherent drive on the cavity may occur, except for the generated parametric drive on the cavity. This corresponds to adding $\varepsilon_a(a^{\dag}e^{-i\omega_{d}t}+ae^{i\omega_{d}t})$ to the drive Hamiltonian $H_{d}$ in Eq.~(\ref{parametric-drive}), where $\varepsilon_a$ is the Rabi frequency related to the additional coherent drive.
Here this additional coherent drive is directly on the cavity, instead of the YIG sphere, so its effect on the magnon mode can be ignored.

When including this additional coherent drive on the cavity, it follows from Eqs.~(\ref{main-Hamiltonian})-(\ref{parametric-drive}) that the total Hamiltonian of the system becomes
\begin{eqnarray}\label{H1}
H&=&\omega_c a^{\dag}a+\omega_m b^{\dag}b+\frac{K}{2}b^{\dag}b^{\dag}bb+g(a^{\dag}b+ab^{\dag})\nonumber\\
   & &+\frac{G}{2}(a^{\dag}a^{\dag}e^{-i\omega_{d}t}+aae^{i\omega_{d}t})
      +\varepsilon_a(a^{\dag}e^{-i\omega_{d}t}+ae^{i\omega_{d}t}).~~~~~~
\end{eqnarray}
Transforming $H$ into the interaction picture via the unitary transformation
$R=\exp(-i\omega_c a^{\dag}at-i\omega_m b^{\dag}bt)$, we can convert the total Hamiltonian $H$ to
\begin{eqnarray}\label{H2}
H' &=&R^\dag H R-iR^\dag \frac{\partial R}{\partial t}  \nonumber\\
    &=&\frac{K}{2}b^{\dag}b^{\dag}bb+g\big[a^{\dag}be^{-i(\omega_{m}-\omega_{c})t}
         +ab^{\dag}e^{i(\omega_{m}-\omega_{c})t}\big]\nonumber\\
    & &+\frac{G}{2}\big[a^{\dag}a^{\dag}e^{-i(\omega_{d}-2\omega_c)t}+aae^{i(\omega_{d}-2\omega_c)t}\big]\nonumber\\
    & &+\varepsilon_a\big[a^{\dag}e^{-i(\omega_{d}-\omega_c)t}
       +ae^{i(\omega_{d}-\omega_c)t}\big].
\end{eqnarray}
To engineer the parity-symmetry-breaking QPT, the parameters of the system are chosen to satisfy $\omega_{d} \approx 2\omega_{c} \approx 2\omega_{m}$ in our scheme. Under this condition, the two-photon processes related to $a^\dag a^\dag$ and $aa$ dominate,
owing to the designed match of the drive-field frequency $\omega_d$ with the two-photon energy (i.e., $\omega_{d} \approx 2\omega_{c}$). In contrast, $\varepsilon_a a^{\dag}e^{-i(\omega_d-\omega_c)t}$ and $\varepsilon_a ae^{i(\omega_d-\omega_c)t}$ then become fast-oscillating terms, due to the large frequency detuning $\omega_d-\omega_c \approx \omega_c$. According to the rotating-wave approximation~\cite{Walls94}, these fast-oscillating terms can be neglected, even if $\varepsilon_a$ is comparable to $G$.

Furthermore, in case these fast-oscillating terms cannot be fully ignored, one can also introduce another coherent drive on the cavity with the same frequency $\omega_d$ but a different phase $\Delta\phi$, $\varepsilon'_a[a^{\dag}e^{-i(\omega_{d}t+\Delta\phi)}+ae^{i(\omega_{d}t+\Delta\phi)}]$, where $\varepsilon'_a$ is the corresponding Rabi frequency. Experimentally, it can be readily implemented by tuning $\varepsilon'_a$ and $\Delta\phi$ via the drive power and a phase shifter, respectively~\cite{Zhang17}. When $\varepsilon'_a$ and $\Delta\phi$ are tuned to have $\varepsilon'_a=\varepsilon_a$ and $\Delta\phi=\pi$, this drive field then cancels the fast-oscillating terms in Eq.~(\ref{H2}) [i.e., cancels $\varepsilon_a(a^{\dag}e^{-i\omega_{d}t}+ae^{i\omega_{d}t})$ in Eq.~(\ref{H1})], leaving only the parametric drive on the cavity.

In our work, we focus on the YIG sphere, which is widely used in experiments on cavity magnonics~\cite{Lachance-Quirion19,Rameshti2021}. Compared with YIG samples of other geometric shapes, the YIG sphere has two advantages. First, the spherical YIG sample avoids the effect of the magnetic dipole-dipole interaction~\cite{Gurevich96,Stancil09}, since the part of the Hamiltonian contributed by the magnetic dipole-dipole interaction becomes constant~\cite{Zhang-China-19}. As shown in Ref.~\cite{Wang18}, the Kerr-nonlinearity Hamiltonian in Eq.~(\ref{Kerr}) is derived for a YIG sphere, which describes the magnon-magnon interaction. When the YIG sample is not spherical, the results deviate, because the contribution of the dipole-dipole interaction to the Hamiltonian is not constant now. Second, the magnon mode in the YIG sphere has a smaller damping rate than the nonspherical YIG sample~\cite{Huebl13,Morris17}. This means that a stronger parametric drive is needed to realize the QPT for a nonspherical YIG sample.

To summarize, we have proposed an experimentally feasible scheme to engineer the nonequilibrium QPT in a parametrically driven cavity magnonic system containing the spin ensemble in a ferrimagnetic YIG sphere. By investigating its steady state in the presence of decoherence, we obtain the steady-state phase diagram of the system. With appropriate parameters, it is found that the cavity magnonic system can undergo both first- and second-order QPTs by just varying the drive-field strength. Experimentally, the CWR with a parametric drive has been achieved~\cite{Yamamoto08,Zhong13,Lin13}, and the strong coupling between magnon and CWR modes is also demonstrated~\cite{Huebl13,Morris17,Hou19,Li19}. Thus, our scheme is implementable in a quantum circuit consisting of a YIG sphere and a CWR using the parametric drive.

\section*{Acknowledgments}

This work is supported by the National Natural Science Foundation of China (Grants No. 11774022, No. U1801661, No. 11934010, and No.~11804074), the Postdoctoral Science Foundation of China (Grant No. 2020M671687), and Zhejiang Province Program for Science and Technology (Grant No. 2020C01019). C.H.L. is supported by Hong Kong General Research Fund (Grant No. 15301717).

\appendix

\section{Master equation of the cavity magnonic system}\label{Appendix-A}

In the presence of decoherence, the evolution of the system can be expressed using the Lindblad master equation~\cite{Walls94}:
\begin{eqnarray}\label{master-equation}
\dot{\rho}&=&i[\rho,H_{\rm tot}]+\frac{\kappa_{r}}{2}\mathcal{D}[a]\rho+\kappa_{\varphi}\mathcal{D}[a^{\dag}a]\rho\nonumber\\
          & &         +\frac{\gamma_{r}}{2}\mathcal{D}[b]\rho+\gamma_{\varphi}\mathcal{D}[b^{\dag}b]\rho,
\end{eqnarray}
where the total Hamiltonian $H_{\rm tot}$ of the system is given in Eq.~(\ref{Hermitian-Hamiltonian}), and the Lindblad dissipators are
\begin{eqnarray}
\mathcal{D}[\mathcal{O}]\rho=(2\mathcal{O}\rho \mathcal{O}^{\dag}
                                -\mathcal{O}^{\dag}\mathcal{O}\rho-\rho \mathcal{O}^{\dag}\mathcal{O}),
\end{eqnarray}
with $\mathcal{O}=a,~a^{\dag}a,~b,$ and $b^{\dag}b$. In Eq.~(\ref{master-equation}), $\kappa_r=1/T^{(a)}_r$ ($\gamma_r=1/T^{(b)}_r$) is the relaxation rate of the cavity (magnon) mode, with $T^{(a)}_r$ ($T^{(b)}_r$) being the relaxation time, and $\kappa_\varphi = 1/T^{(a)}_\varphi$ ($\gamma_\varphi = 1/T^{(b)}_\varphi$) is the pure dephasing rate of the cavity (magnon) mode, with $T^{(a)}_\varphi$ ($T^{(b)}_\varphi$) being the dephasing time.

With the master equation in Eq.~(\ref{master-equation}), we can obtain the equation of motion for the expectation value $\langle \mathcal{O}\rangle$ via the relation $\langle \dot{\mathcal{O}}\rangle={\rm Tr}(\dot{\rho}\mathcal{O})$,
\begin{eqnarray}\label{master-expected-value}
\langle\dot{a}\rangle&=&-i(\Delta_{c}-i\kappa)\langle a\rangle -ig\langle b\rangle-iG\langle a^{\dag}\rangle,\nonumber\\
\langle\dot{b}\rangle&=&-i(\Delta_{m}-i\gamma)\langle b\rangle
                        -iK \langle b^{\dag}b b\rangle-ig\langle a\rangle,
\end{eqnarray}
and
\begin{eqnarray}\label{number-operator}
\frac{d}{dt}\langle a^{\dag}a\rangle&=&-\kappa_{r}\langle a^{\dag}a\rangle
                                        -ig(\langle a^{\dag}b\rangle-\langle ab^{\dag}\rangle)
                                       -iG(\langle a^{\dag}a^{\dag}\rangle-\langle aa\rangle),\nonumber\\
\frac{d}{dt}\langle b^{\dag}b\rangle&=&-\gamma_{r}\langle b^{\dag}b\rangle
                                        -ig(\langle ab^{\dag}\rangle-\langle a^{\dag}b\rangle),
\end{eqnarray}
where $\kappa =\frac{1}{2}\kappa_r+\kappa_\varphi$ is the decoherence rate of the cavity mode, and $\gamma=\frac{1}{2}\gamma_r+\gamma_\varphi$ is the decoherence rate of the magnon mode~\cite{Wang09}. This is similar to the decoherence rate of a two-level system (see, e.g., Ref.~\cite{You07}). Under the mean-field approximation $\langle b^{\dag}b b\rangle \approx \langle b^{\dag}\rangle\langle b\rangle\langle b\rangle$, Eq.~(\ref{master-expected-value}) is reduced to Eq.~(\ref{expected-value}) in the main text.

As discussed in Sec.~\ref{conclusions}, the proposed QPT can be implemented in a CWR embedded with a YIG sphere. For the CWR mode, its relaxation rate is 30 times larger than the pure dephasing rate (i.e., $\kappa_r/\kappa_\varphi\approx 30$) at tens of milli-Kelvin temperatures~\cite{Wang09}. When the temperatures range from 0 to 1~K, the relaxation of the magnon mode in YIG dominates ($\gamma_r\sim 1$~MHz) and the pure dephasing can be ignored~\cite{Tabuchi14}. Thus, at low temperatures, we can use $\kappa=\frac{1}{2}\kappa_r+\kappa_\varphi\approx \frac{1}{2}\kappa_r$ and $\gamma=\frac{1}{2}\gamma_r+\gamma_\varphi \approx \frac{1}{2}\gamma_r$ in Eq.~(\ref{master-expected-value}).

\section{Critical exponent of the mean magnon number}\label{Appendix-B}

Around the critical point $G=G_{\rm c2}$ given in Eq.~(\ref{Gc2}), the variation of the mean magnon number is
\begin{eqnarray}\label{eq-A1}
\delta n &=& \langle b^{\dag}b\rangle-\langle b^{\dag}b\rangle|_{G=G_{\rm c2}}\nonumber\\
                               &=&\frac{1}{K}\Big[-\Delta'_{m} + \sqrt{\eta^{2}G^{2}-{\gamma'}^{2}}\Big]\nonumber\\
         &=&\frac{1}{K}\Big[-(\Delta_{m}-\eta\Delta_{c}) + \sqrt{\eta^{2}G^{2}-(\gamma+\eta\kappa)^{2}}\Big].
\end{eqnarray}
When keeping terms up to the first order $\sim G-G_{\rm c2}$, $\eta=g^{2}/(\Delta_{c}^{2}+\kappa^{2}-G^{2})$ can be approximatively expressed as
\begin{eqnarray}\label{eq-A2}
\eta &=& \frac{g^{2}}{\Delta_{c}^{2}+\kappa^{2}-[G_{\rm c2}+(G-G_{\rm c2})]^{2}}\nonumber\\
     &\approx& \eta_{c}+\lambda_{1}(G-G_{\rm c2}),
\end{eqnarray}
where $\eta_{c}=g^{2}/(\Delta_{c}^{2}+\kappa^{2}-G_{\rm c2}^{2})$ and $\lambda_{1}=2\eta_{c}^{2}G_{\rm c2}/g^{2}$. With the relation in the above equation, we obtain
\begin{eqnarray}\label{eq-A3}
\Delta_{m}-\eta\Delta_{c}\approx\Delta_{m}-\eta_{c}\Delta_{c}-\lambda_{1}\Delta_{c}(G-G_{\rm c2}),
\end{eqnarray}
\begin{eqnarray}\label{eq-A4}
\sqrt{\eta^{2}G^{2}-(\gamma+\eta\kappa)^{2}} \approx \sqrt{\eta_{c}^{2}G_{\rm c2}^{2}-(\gamma+\eta_{c}\kappa)^{2}}
                                             +\lambda_{2}(G-G_{\rm c2}),\nonumber\\
\end{eqnarray}
where
\begin{eqnarray}\label{eq-A5}
\lambda_{2}=\frac{\eta_{c}G_{\rm c2}(\eta_{c}+\lambda_{1}G_{\rm c2})-\lambda_{1}\kappa(\gamma+\eta_{c}\kappa)}
                                 {\sqrt{\eta_{c}^{2}G_{\rm c2}^{2}-(\gamma+\eta_{c}\kappa)^{2}}}.
\end{eqnarray}
Substituting the expressions in Eqs.~(\ref{eq-A3}) and (\ref{eq-A4}) into Eq.~(\ref{eq-A1}), we obtain
\begin{eqnarray}\label{eq-A6}
\lim_{G\rightarrow G_{\rm c2}}\delta n &=& \frac{\lambda_{1}\Delta_{c}+\lambda_{2}}{K}(G-G_{\rm c2})\sim(G-G_{\rm c2})^{\nu},
\end{eqnarray}
which has the critical exponent $\nu=1$.

\section{Dynamics of the correlated fluctuations}\label{Appendix-C}

From Eq.~(\ref{fluctuations}), the equations of motion for the correlated fluctuations are obtained as
\begin{eqnarray}\label{eq-B1}
\frac{d}{dt}(\delta a\delta b^{\dag})&=&-i[(\Delta_{c}-\widetilde{\Delta}_{m})-i(\kappa+\gamma)] \delta a\delta b^{\dag}
                                      -ig(\delta b^{\dag}\delta b-\delta a^{\dag}\delta a)\nonumber\\
                        & &   -iG\delta a^{\dag}\delta b^{\dag}+iF^{*}\delta a \delta b
                            +(\sqrt{2\kappa}a_{\rm in}\delta b^{\dag}+\sqrt{2\gamma}\delta a b_{\rm in}^{\dag}),\nonumber\\
\frac{d}{dt}(\delta a\delta b)&=&-i[(\Delta_{c}+\widetilde{\Delta}_{m})-i(\kappa+\gamma)]\delta a\delta b
                                 -ig(\delta a\delta a+\delta b\delta b)\nonumber\\
                              & & -iG\delta a^{\dag}\delta b-iF\delta a\delta b^{\dag}
                                   +(\sqrt{2\kappa}a_{\rm{in}}\delta b+\sqrt{2\gamma}\delta a b_{\rm in}),\nonumber\\
\frac{d}{dt}(\delta a\delta a)&=&-2i(\Delta_{c}-i\kappa)\delta a\delta a-2ig\delta a\delta b
                                  -iG(2\delta a^{\dag}\delta a+1)\nonumber\\
                              & & +\sqrt{2\kappa}(a_{\rm{in}}\delta a+\delta a a_{\rm in}),\nonumber\\
\frac{d}{dt}(\delta b\delta b)&=&-2i(\widetilde{\Delta}_{m}-i\gamma) \delta b\delta b-2ig\delta a\delta b
                                   -iF(2\delta b^{\dag}\delta b+1)\nonumber\\
                              & &  +\sqrt{2\gamma}(b_{\rm{in}}\delta b+\delta b b_{\rm{in}}),\nonumber\\
\frac{d}{dt}(\delta a^{\dag}\delta a)&=&-2\kappa \delta a^{\dag}\delta a
                                      -ig(\delta a^{\dag}\delta b-\delta a \delta b^{\dag})
                                      -iG\delta a^{\dag}\delta a^{\dag}\nonumber\\
                       & &+iG\delta a \delta a+\sqrt{2\kappa}(\delta a^{\dag}a_{\rm{in}}+a_{\rm{in}}^{\dag}\delta a),\nonumber\\
\frac{d}{dt}(\delta b^{\dag}\delta b)&=&-2\gamma \delta b^{\dag}\delta b-ig(\delta a\delta b^{\dag}-\delta a^{\dag}\delta b)
                                         -iF\delta b^{\dag}\delta b^{\dag}\nonumber\\
                                     & &+iF^{*}\delta b \delta b+\sqrt{2\gamma}(\delta b^{\dag}b_{\rm{in}}
                                           +b_{\rm{in}}^{\dag}\delta b),
\end{eqnarray}
where $\widetilde{\Delta}_{m}=\Delta_{m}+2K\langle b^{\dag}\rangle\langle b\rangle$ and $F=K\langle b\rangle^{2}$. Then, it follows from Eq.~(\ref{eq-B1}) that the expectation values $A_{1}\equiv\langle \delta a\delta b^{\dag}\rangle$, $A_{2}\equiv\langle \delta a\delta b \rangle$, $A_{3}\equiv\langle \delta a\delta a \rangle$, $A_{4}\equiv\langle \delta b\delta b \rangle$, $A_{5}\equiv\langle \delta a^{\dag}\delta a \rangle$ and $A_{6}\equiv\langle \delta b^{\dag}\delta b \rangle$ satisfy
\begin{eqnarray}\label{eq-B2}
\dot{A}_{1}&=&-i[(\Delta_{c}-\widetilde{\Delta}_{m})-i(\kappa+\gamma)] A_{1}-ig(A_{6}-A_{5})-iGA^*_{2}+iF^{*}A_{2},\nonumber\\
\dot{A}_{2}&=&-i[(\Delta_{c}+\widetilde{\Delta}_{m})-i(\kappa+\gamma)]A_{2}-ig(A_{3}+A_{4})-iGA^*_{1}-iFA_{1},\nonumber\\
\dot{A}_{3}&=&-2i(\Delta_{c}-i\kappa)A_{3}-2igA_{2}-iG(2A_{5}+1),\nonumber\\
\dot{A}_{4}&=&-2i(\widetilde{\Delta}_{m}-i\gamma) A_{4}-2igA_{2}-iF(2A_{6}+1),\nonumber\\
\dot{A}_{5}&=&-2\kappa A_{5}-ig(A^*_{1}-A_{1})-iGA^*_{3}+iGA_{3},\nonumber\\
\dot{A}_{6}&=&-2\gamma A_{6}-ig(A_{1}-A^*_{1})-iFA^*_{4}+iF^{*}A_{4}.
\end{eqnarray}
At the steady state, $\dot{A}_{j}=0$, $j=1$ to 6. We can investigate the expectation values of the correlated fluctuations $\delta a^{\dag}\delta a$ and $\delta b^{\dag}\delta b$ for both cavity photons and magnons (i.e., $A_{5}$ and $A_{6}$) by numerically solving Eq.~(\ref{eq-B2}).


\begin{thebibliography}{99}

\bibitem{Xiang13}
Z. L. Xiang, S. Ashhab, J. Q. You, and F. Nori,
Hybrid quantum circuits: Superconducting circuits interacting with other quantum systems,
Rev. Mod. Phys. \textbf{85}, 623 (2013).

\bibitem{Forn-Diaz19}
P. Forn-D\'{\i}az, L. Lamata, E. Rico, J. Kono, and E. Solano,
Ultrastrong coupling regimes of light-matter interaction,
Rev. Mod. Phys. \textbf{91}, 025005 (2019).

\bibitem{Kockum2019}
A. F. Kockum, A. Miranowicz, S. D. Liberato, S. Savasta and F. Nori, Ultrastrong coupling between light and matter, Nat. Rev. Phys. \textbf{1}, 19 (2019).

\bibitem{Dicke54}
R. H. Dicke,
Coherence in spontaneous radiation processes,
Phys. Rev. \textbf{93}, 99 (1954).

\bibitem{Emary-2003-PRE}
C. Emary and T. Brandes,
Chaos and the quantum phase transition in the Dicke model,
Phys. Rev. E \textbf{67}, 066203 (2003).

\bibitem{Emary-2003-PRL}
C. Emary and T. Brandes,
Quantum Chaos Triggered by Precursors of a Quantum Phase Transition: The Dicke Model,
Phys. Rev. Lett. \textbf{90}, 044101 (2003).

\bibitem{Hepp73}
K. Hepp and E. H. Lieb,
On the superradiant phase transition for molecules in a quantized radiation field: The Dicke maser model,
Ann. Phys. (NY) \textbf{76}, 360 (1973).

\bibitem{Wang73}
Y. K. Wang and F. T. Hioe,
Phase transition in the dicke model of superradiance,
Phys. Rev. A \textbf{7}, 831 (1973).

\bibitem{Dimer07}
F. Dimer, B. Estienne, A. S. Parkins, and H. J. Carmichael,
Proposed realization of the Dicke-model quantum phase transition in an optical cavity QED system,
Phys. Rev. A \textbf{75}, 013804 (2007).

\bibitem{Nagy10}
D. Nagy, G. K\'{o}nya, G. Szirmai, and P. Domokos,
Dicke-Model Phase Transition in the Quantum Motion of a Bose-Einstein Condensate in an Optical Cavity,
Phys. Rev. Lett. \textbf{104}, 130401 (2010).

\bibitem{You20}
G. Q. Zhang, Z. Chen, and J. Q. You,
Experimentally accessible quantum phase transition in a non-Hermitian Tavis-Cummings model engineered with two drive fields,
Phys. Rev. A \textbf{102}, 032202 (2020).

\bibitem{Zhu20}
C. J. Zhu, L. L. Ping, Y. P. Yang, and G. S. Agarwal,
Squeezed Light Induced Symmetry Breaking Superradiant Phase Transition,
Phys. Rev. Lett. \textbf{124}, 073602 (2020).

\bibitem{Baumann10}
K. Baumann, C. Guerlin, F. Brennecke, and T. Esslinger,
Dicke quantum phase transition with a superfluid gas in an optical cavity,
Nature (London) \textbf{464}, 1301 (2010).

\bibitem{Baden14}
M. P. Baden, K. J. Arnold, A. L. Grimsmo, S. Parkins, and M. D. Barrett,
Realization of the Dicke Model Using Cavity-Assisted Raman Transitions,
Phys. Rev. Lett. \textbf{113}, 020408 (2014).

\bibitem{Tabuchi14}
Y. Tabuchi, S. Ishino, T. Ishikawa, R. Yamazaki, K. Usami, and Y. Nakamura,
Hybridizing Ferromagnetic Magnons and Microwave Photons in the Quantum Limit,
Phys. Rev. Lett. \textbf{113}, 083603 (2014).

\bibitem{Zhang14}
X. Zhang, C. L. Zou, L. Jiang, and H. X. Tang, Strongly Coupled Magnons and Cavity Microwave Photons, Phys. Rev. Lett. \textbf{113}, 156401 (2014).

\bibitem{SoykalPRL10}
\"{O}. O. Soykal and M. E. Flatt\'{e},
Strong Field Interactions between a Nanomagnet and a Photonic Cavity,
Phys. Rev. Lett. \textbf{104}, 077202 (2010).

\bibitem{Zhang15-1}
D. Zhang, X. M. Wang, T. F. Li, X. Q. Luo, W. Wu, F. Nori, and J. Q. You,
Cavity quantum electrodynamics with ferromagnetic magnons in a small yttrium-iron-garnet sphere,
npj Quantum Information \textbf{1}, 15014 (2015).

\bibitem{Cao15}
Y. Cao, P. Yan, H. Huebl, S. T. B. Goennenwein, and G. E. W. Bauer,
Exchange magnon-polaritons in microwave cavities,
Phys. Rev. B \textbf{91}, 094423 (2015).

\bibitem{Bai15}
L. Bai, M. Harder, Y. P. Chen, X. Fan, J. Q. Xiao, and C.-M. Hu,
Spin Pumping in Electrodynamically Coupled Magnon-Photon Systems,
Phys. Rev. Lett. \textbf{114}, 227201 (2015).

\bibitem{Tabuchi15}
Y. Tabuchi, S. Ishino, A. Noguchi, T. Ishikawa, R. Yamazaki, K. Usami, and Y. Nakamura,
Coherent coupling between a ferromagnetic magnon and a superconducting qubit,
Science \textbf{349}, 405 (2015).

\bibitem{Li18}
J. Li, S.-Y. Zhu, and G. S. Agarwal,
Magnon-Photon-Phonon Entanglement in Cavity Magnomechanics,
Phys. Rev. Lett. \textbf{121}, 203601 (2018).

\bibitem{Shim20}
J. Shim, S.-J. Kim, S. K. Kim, and K.-J. Lee,
Enhanced Magnon-Photon Coupling at the Angular Momentum Compensation Point of Ferrimagnets,
Phys. Rev. Lett. \textbf{125}, 027205 (2020).

\bibitem{Lachance-Quirion19}
D. Lachance-Quirion, Y. Tabuchi, A. Gloppe, K. Usami, and Y. Nakamura,
Hybrid quantum systems based on magnonics,
Appl. Phys. Express \textbf{12}, 070101 (2019).

\bibitem{Rameshti2021}
B. Z. Rameshti, S. V. Kusminskiy, J. A. Haigh, K. Usami, D. Lachance-Quirion, Y. Nakamura, C. M. Hu, H. X. Tang, G. E. W. Bauer, and Y. M. Blanter,
Cavity Magnonics,
arXiv:2106.09312.

\bibitem{Zhang15-2}
X. Zhang, C.-L. Zou, N. Zhu, F. Marquardt, L. Jiang, and H. X. Tang,
Magnon dark modes and gradient memory,
Nat. Commun. \textbf{6}, 8914 (2015).

\bibitem{Zhang17}
D. Zhang, X. Q. Luo, Y. P. Wang, T. F. Li, and J. Q. You,
Observation of the exceptional point in cavity magnon-polaritons,
Nat. Commun. \textbf{8}, 1368 (2017).

\bibitem{Zhang19}
X. Zhang, K. Ding, X. Zhou, J. Xu, and D. Jin,
Experimental Observation of an Exceptional Surface in Synthetic Dimensions with Magnon Polaritons,
Phys. Rev. Lett. \textbf{123}, 237202 (2019).

\bibitem{Zhao20}
J. Zhao, Y. Liu, L. Wu, C. K. Duan, Y. Liu, and J. Du,
Observation of Anti-$\mathcal{PT}$-Symmetry Phase Transition in the Magnon-Cavity-Magnon Coupled System,
Phys. Rev. Appl. \textbf{13}, 014053 (2020).

\bibitem{You19}
G. Q. Zhang and J. Q. You,
Higher-order exceptional point in a cavity magnonics system,
Phys. Rev. B \textbf{99}, 054404 (2019).

\bibitem{Cao19}
Y. Cao and P. Yan,
Exceptional magnetic sensitivity of $\mathcal{PT}$-symmetric cavity magnon polaritons,
Phys. Rev. B \textbf{99}, 214415 (2019).

\bibitem{Harder18}
M. Harder, Y. Yang, B. M. Yao, C. H. Yu, J. W. Rao, Y. S. Gui, R. L. Stamps, and C. M. Hu,
Level Attraction Due to Dissipative Magnon-Photon Coupling,
Phys. Rev. Lett. \textbf{121}, 137203 (2018).

\bibitem{Grigoryan18}
V. L. Grigoryan, K. Shen, and K. Xia,
Synchronized spin-photon coupling in a microwave cavity,
Phys. Rev. B. \textbf{98}, 024406 (2018).

\bibitem{Wang19}
Y. P. Wang, J. W. Rao, Y. Yang, P. C. Xu, Y. S. Gui, B. M. Yao, J. Q. You, and C.-M. Hu,
Nonreciprocity and Unidirectional Invisibility in Cavity Magnonics,
Phys. Rev. Lett. \textbf{123}, 127202 (2019).

\bibitem{Yu19}
W. Yu, J. Wang, H. Y. Yuan, and J. Xiao,
Prediction of Attractive Level Crossing via a Dissipative Mode,
Phys. Rev. Lett. \textbf{123}, 227201 (2019).

\bibitem{Rao21}
J. W. Rao, P. C. Xu, Y. S. Gui, Y. P. Wang, Y. Yang, B. Yao, J. Dietrich, G. E. Bridges, X. L. Fan, D. S. Xue, and C. M. Hu,
Interferometric control of magnon-induced nearly perfect absorption in cavity magnonics,
Nat. Commun. \textbf{12}, 1933 (2021).

\bibitem{Yuan-20-1}
H. Y. Yuan, S. Zheng, Q. Y. He, J. Xiao, and R. A. Duine,
Unconventional magnon excitation by off-resonant microwaves,
Phys. Rev. B \textbf{103}, 134409 (2021)

\bibitem{Yang21}
Z. B. Yang, X. D. Liu, X. Y. Yin, Y. Ming, H. Y. Liu, and R. C. Yang,
Controlling Stationary One-Way Quantum Steering in Cavity Magnonics,
Phys. Rev. Applied \textbf{15}, 024042 (2021).

\bibitem{Zhang-China-19}
G. Q. Zhang, Y. P. Wang, and J. Q. You,
Theory of the magnon Kerr effect in cavity magnonics,
Sci. China-Phys. Mech. Astron. \textbf{62}, 987511 (2019).

\bibitem{Wang16}
Y. P. Wang, G. Q. Zhang, D. Zhang, X. Q. Luo, W. Xiong, S. P. Wang, T. F. Li, C. M. Hu, and J. Q. You,
Magnon Kerr effect in a strongly coupled cavity-magnon system,
Phys. Rev. B \textbf{94}, 224410 (2016).

\bibitem{Wang18}
Y. P. Wang, G. Q. Zhang, D. Zhang, T. F. Li, C. M. Hu, and J. Q. You,
Bistability of Cavity Magnon-Polaritons,
Phys. Rev. Lett. \textbf{120}, 057202 (2018).

\bibitem{Nair20}
J. M. P. Nair, Z. Zhang, M. O. Scully, and G. S. Agarwal,
Nonlinear spin currents,
Phys. Rev. B \textbf{102}, 104415 (2020).

\bibitem{Bi21}
M. X. Bi, X. H. Yan, Y. Zhang, and Y. Xiao,
Tristability of cavity magnon polaritons,
Phys. Rev. B \textbf{103}, 104411 (2021).

\bibitem{Kong19}
C. Kong, H. Xiong, and Y. Wu,
Magnon-Induced Nonreciprocity Based on the Magnon Kerr Effect,
Phys. Rev. Appl. \textbf{12}, 034001 (2019).

\bibitem{Scully19}
Z. Zhang, M. O. Scully, and G. S. Agarwal,
Quantum entanglement between two magnon modes via Kerr nonlinearity driven far from equilibrium,
Phys. Rev. Research \textbf{1}, 023021 (2019).

\bibitem{Gurevich96}
A. G. Gurevich and G. A. Melkov,
\emph{Magnetization Oscillations and Waves} (CRC Press, Boca Raton, 1996).

\bibitem{Stancil09}
D. D. Stancil and A. Prabhakar,
\emph{Spin Waves} (Springer, Berlin, 2009).

\bibitem{Sachdev11}
S. Sachdev,
\emph{Quantum Phase Transitions} (Cambridge University Press, Cambridge, 1999).

\bibitem{Lambert04}
N. Lambert, C. Emary, and T. Brandes,
Entanglement and the Phase Transition in Single-Mode Superradiance,
Phys. Rev. Lett. \textbf{92}, 073602 (2004).

\bibitem{Vidal03}
G. Vidal, J. I. Latorre, E. Rico, and A. Kitaev,
Entanglement in Quantum Critical Phenomena,
Phys. Rev. Lett. \textbf{90}, 227902 (2003).

\bibitem{Osborne02}
T. J. Osborne and M. A. Nielsen,
Entanglement in a simple quantum phase transition,
Phys. Rev. A \textbf{66}, 032110 (2002).

\bibitem{Walls94}
D. F. Walls and G. J. Milburn,
\emph{Quantum Optics} (Springer, Berlin, 2007).

\bibitem{Hyde18}
P. Hyde, B. M. Yao, Y. S. Gui, G. Q. Zhang, J. Q. You, and C. M. Hu,
Direct measurement of foldover in cavity magnon-polariton systems,
Phys. Rev. B \textbf{98}, 174423 (2018).

\bibitem{Zhang20}
X. H. H. Zhang and H. U. Baranger,
Driven-dissipative phase transition in a Kerr oscillator: From semiclassical $\mathcal{PT}$ symmetry to quantum fluctuations,
Phys. Rev. A \textbf{103}, 033711 (2021).

\bibitem{Gradshteyn80}
I. S. Gradshteyn and I. M. Ryzhik,
\emph{Table of Integrals, Series and Products} (Academic, Orlando, 1980).

\bibitem{Nagy11}
D. Nagy, G. Szirmai, and P. Domokos,
Critical exponent of a quantum-noise-driven phase transition: The open-system Dicke model,
Phys. Rev. A \textbf{84}, 043637 (2011).

\bibitem{Zheng11}
S. B. Zheng,
Dicke-like quantum phase transition and vacuum entanglement with two coupled atomic ensembles,
Phys. Rev. A \textbf{84}, 033817 (2011).

\bibitem{Huebl13}
H. Huebl, C. W. Zollitsch, J. Lotze, F. Hocke, M. Greifenstein, A. Marx, R. Gross, and S. T. B. Goennenwein,
High Cooperativity in Coupled Microwave Resonator Ferrimagnetic Insulator Hybrids,
Phys. Rev. Lett. \textbf{111}, 127003 (2013).

\bibitem{Morris17}
R. G. E. Morris, A. F. van Loo, S. Kosen, and A. D. Karenowska,
Strong coupling of magnons in a YIG sphere to photons in a planar superconducting resonator in the quantum limit,
Sci. Rep. \textbf{7}, 11511 (2017).

\bibitem{Hou19}
J. T. Hou and L. Liu,
Strong Coupling between Microwave Photons and Nanomagnet Magnons,
Phys. Rev. Lett. \textbf{123}, 107702 (2019).

\bibitem{Li19}
Y. Li, T. Polakovic, Y.-L.Wang, J. Xu, S. Lendinez, Z. Zhang, J. Ding, T. Khaire, H. Saglam, R. Divan, J. Pearson, W. K. Kwok, Z. Xiao, V. Novosad, A. Hoffmann, and W. Zhang,
Strong Coupling between Magnons and Microwave Photons in On-Chip Ferromagnet-Superconductor Thin-Film Devices,
Phys. Rev. Lett. \textbf{123}, 107701 (2019).

\bibitem{Yamamoto08}
T. Yamamoto, K. Inomata, M. Watanabe, K. Matsuba, T. Miyazaki, W. D. Oliver, Y. Nakamura, and J. S. Tsai,
Flux-driven Josephson parametric amplifier,
Appl. Phys. Lett. \textbf{93}, 042510 (2008).

\bibitem{Zhong13}
L. Zhong, E. P. Menzel, R. D. Candia, P. Eder, M. Ihmig, A. Baust, M. Haeberlein, E. Hoffmann, K. Inomata, T.
Yamamoto, Y. Nakamura, E. Solano, F. Deppe, A. Marx, and R. Gross,
Squeezing with a flux-driven Josephson parametric amplifer,
New J. Phys. \textbf{15}, 125013 (2013).

\bibitem{Lin13}
Z. R. Lin, K. Inomata, W. D. Oliver, K. Koshino, Y. Nakamura, J. S. Tsai, and T. Yamamoto,
Single-shot readout of a superconducting flux qubit with a flux-driven Josephson parametric amplifier,
Appl. Phys. Lett. \textbf{103}, 132602 (2013).

\bibitem{Krantz16}
P. Krantz, A. Bengtsson, M. Simoen, S. Gustavsson, V. Shumeiko, W. D. Oliver, C. M. Wilson, P. Delsing, and J. Bylander, Single-shot read-out of a superconducting qubit using a Josephson parametric oscillator,
Nat. Commun. \textbf{7}, 11417 (2016).

\bibitem{Wang09}
H. Wang, M. Hofheinz, M. Ansmann, R. C. Bialczak, E. Lucero, M. Neeley, A. D. O'Connell, D. Sank, M. Weides, J. Wenner, A. N. Cleland, and J. M. Martinis,
Decoherence Dynamics of Complex Photon States in a Superconducting Circuit,
Phys. Rev. Lett. \textbf{103}, 200404 (2009).

\bibitem{You07}
J. Q. You, X. Hu, S. Ashhab, and F. Nori,
Low-decoherence flux qubit,
Phys. Rev. B \textbf{75}, 140515 (2007).

\end{thebibliography}
\end{document}